# Trace Identities for Skew-Symmetric Matrices

M. I. Krivoruchenko[1, 2, 3]

[1]Theoretical Physics Division, Institute for Theoretical and Experimental Physics, Moscow, Russia
[2]Department of Nano-, Bio-, Information and Cognitive Technologies, Moscow Institute of Physics and Technology, Dolgoprudny, Russia
[3]Bogoliubov Laboratory of Theoretical Physics, Joint Institute for Nuclear Research, Dubna, Russia

**Email address:**
mikhail.krivoruchenko@itep.ru

**Abstract:** We derive an expression for the product of the Pfaffians of two skew-symmetric matrices $A$ and $B$ as a sum of products of the traces of powers of $AB$ and an expression for the inverse matrix $A^{-1}$, or equivalently $B^{-1}$, as a finite-order polynomial of $AB$ with coefficients depending on the traces of powers of $AB$.

## 1. Introduction

The Cayley-Hamilton theorem [1] states that for any $n \times n$ matrix $C$ there exists a polynomial $P_n(z) = a_0 + a_1 z + \ldots + a_n z^n$ that vanishes when $z$ is replaced by $C$: $P_n(C) = 0$. The coefficients $a_m$ are expressed in terms of the traces of powers of $C$; these coefficients can be calculated recursively [2]. The eigenvalues of $C$ are roots of the characteristic equation $P_n(\lambda) = 0$.

In Ref. [3] an explicit expression is found for $a_m$ in terms of the complete Bell polynomials, whose arguments are simply related to the traces of powers of $C$. An equivalent closed expression for $a_m$ in terms of the determinant of a $n \times n$ matrix, whose elements are the traces of powers of $C$, is reported in Ref. [4].

Trace of odd powers of a skew-symmetric matrix $A$ is zero, so odd powers of $A$ do not enter the characteristic polynomial (i.e., $P_{2n}(A) = P_{2n}(-A)$). Accordingly, one can expect that the trace identities for skew-symmetric matrices are simpler than prescribed by the general formulas [3, 4].

In quantum field theory, the Pfaffian arises when integrating Gaussian function of Grassmann variables. The overlap integral between two fermion quasiparticle states is expressed through the Pfaffian [5, 6]. Fermions, as a rule, carry intrinsic quantum numbers besides the spin, so in the applications to physics matrices are often multi-dimensional. The trace of matrices with multi-dimensional indices is the direct sum of the traces over the intrinsic quantum numbers. Presentation of the determinant and the Pfaffian in terms of the trace of the matrix products put the intrinsic symmetries under control at intermediate stages of computing.

The Pfaffian also occurs in the problem of enumeration of dimers (see, e.g., [7]), which is the important problem of chemistry and statistical physics.

In this paper, we report trace identities for the pair of skew-symmetric matrices. In the next Sect., new proofs are given for the trace representations of the determinant and the characteristic polynomial of arbitrary matrix. In Sect. 3, trace identities for skew-symmetric matrices are derived. First, we show that the Pfaffian of a $2n \times 2n$ skew-symmetric matrix $A$ admits decomposition over the traces of powers of $A$ up to the $n$-th degree. Second, we show that the inverse of $A$ is a polynomial of the degree $n-1$ of the product of $A$ and an auxiliary skew-symmetric matrix $B$. The expansion coefficients depend on the traces of powers of $AB$.

## 2. Trace Representation for Characteristic Polynomial of General Matrices

We begin by recalling the definitions (see, e.g., [1]):

*Definition.* A determinant of a $n \times n$ matrix $C$ is a number calculated from

$$\det(C) = \frac{1}{n!} \varepsilon_{i_1 i_2 \ldots i_n} \varepsilon^{j_1 j_2 \ldots j_n} C^{i_1}_{j_1} C^{i_2}_{j_2} \ldots C^{i_n}_{j_n}, \qquad (1)$$

where $1 \leq i_l, j_l \leq n$ for $l = 1, 2, \ldots, n$ and $\varepsilon_{i_1 i_2 \ldots i_n} = \varepsilon^{i_1 i_2 \ldots i_n}$ is the Levi-Civita symbol of the dimension $n$.

*Definition.* A Pfaffian of a $2n \times 2n$ skew-symmetric matrix $A$ is a number calculated from

$$\mathrm{pf}(A) = \frac{1}{2^n n!} \varepsilon_{i_1 j_1 i_2 j_2 \ldots i_n j_n} A^{i_1 j_1} A^{i_2 j_2} \ldots A^{i_n j_n}, \tag{2}$$

where $1 \leq i_l, j_l \leq 2n$ for $l = 1, 2, \ldots, n$ and $\varepsilon_{i_1 j_1 i_2 j_2 \ldots i_n j_n}$ is the Levi-Civita symbol of the dimension $2n$.

The results of Ref. [3] are stated as Lemmas 1 and 2:

*Lemma 1.* Let $C$ be a $n \times n$ matrix, then

$$\det(C) = \sum_{k_1, k_2, \ldots, k_n} \prod_{l=1}^{n} (-1)^{k_l + 1} \frac{\mathrm{tr}(C^l)^{k_l}}{k_l! l^{k_l}}, \tag{3}$$

where the sum runs over sets of non-negative integers $(k_1, k_2, \ldots, k_n)$ satisfying the linear Diophantine equation

$$\sum_{l=1}^{n} l k_l = n. \tag{4}$$

Reference [3] gives a combinatorial proof of the equation. Here we give an alternative

*Proof.* Since the determinant function is homogeneous of degree $n$ in the entries, we write

$$\det(C) = \frac{1}{2\pi i} \oint \frac{dz}{z^{n+1}} \det(I + zC),$$

where $I$ is the unit matrix. We take into account that in the expansion of determinant of the sum of two matrices, after the integration, the term proportional to $z^n$ survives only. The integration is carried out counterclockwise in a closed contour around $z = 0$. The determinant can be represented in the form of the exponent of a trace. In the expansion of the logarithm of $I + zC$, the terms of the degree $z^{n+1}$ and higher, after the expansion of the exponent in the neighborhood of $z = 0$, give vanishing contributions to the contour integral. We can write

$$\det(C) = \frac{1}{2\pi i} \oint \frac{dz}{z^{n+1}} \exp\left(\mathrm{tr} \ln(I + zC)\right)$$

$$= \frac{1}{2\pi i} \oint \frac{dz}{z^{n+1}} \exp\left(\mathrm{tr} \sum_{l=1}^{n} (-1)^{l+1} \frac{z^l C^l}{l}\right).$$

The exponent of the sum is represented in the form of the product of the exponents. The exponents can be expanded in infinite series around $z = 0$. We reorder the terms and obtain

$$\det(C) = \frac{1}{2\pi i} \oint \frac{dz}{z^{n+1}} \sum_{k_1, k_2, \ldots, k_n} \prod_{l=1}^{n} \frac{1}{k_l!} \left((-1)^{l+1} \mathrm{tr}\left(\frac{z^l C^l}{l}\right)\right)^{k_l}$$

$$= \frac{1}{2\pi i} \oint \frac{dz}{z^{n+1}} \sum_{k_1, k_2, \ldots, k_n} \prod_{l=1}^{n} (-1)^{l k_l + k_l} \frac{z^{l k_l} \mathrm{tr}(C^l)^{k_l}}{k_l! l^{k_l}}$$

$$= \frac{1}{2\pi i} \oint \frac{dz}{z^{n+1}} \sum_{k_1, k_2, \ldots, k_n} z^{k_1 + 2k_2 + \ldots + n k_n} \prod_{l=1}^{n} (-1)^{l k_l + k_l} \frac{\mathrm{tr}(C^l)^{k_l}}{k_l! l^{k_l}}.$$

From the representation of the last line it is clear that the integral is different from zero only if the condition (4) holds true. By taking the equation

$$\prod_{l=1}^{n} 2^{l k_l} = 2^{\sum_{l=1}^{n} l k_l} = 2^n$$

into account and performing the contour integral we arrive at Eq. (3). ∎

*Remark 1.* Calculation of the determinant based on the definition requires the $n!$ operations. The number of terms, $v(n)$, entering the expansion (3), grows with $n$ sub-exponentially [8, 9]. $v(n)$ is equal to the number of partitions of $n$; it can be found from the recursion [10]

$$v(m) = \frac{1}{m} \sum_{l=1}^{n} l \sum_{k=1}^{[m/l]} v(m - lk),$$

with the initial conditions $v(m) = 0$ for $m < 0$ and $v(0) = 1$. Here, $[x]$ is the integer part of $x$. Using the Gaussian elimination, the determinant is computed in $O(n^3)$ operations.

*Definition.* Complete Bell polynomials, $B_n(x_1, x_2, \ldots, x_n)$, are defined by the expansion [11]

$$\exp(\sum_{l=1}^{\infty} x_l \frac{z^l}{l!}) = \sum_{n=0}^{\infty} \frac{z^n}{n!} B_n(x_1, \ldots, x_n).$$

*Remark 2.* The right-hand side of Eq. (3) can be recognized, up to a coefficient, as the complete Bell polynomial of $n$ arguments $x_l = -(l-1)! \operatorname{tr}(C^l)$:

$$\det(C) = \frac{(-1)^n}{n!} B_n(x_1, \ldots, x_n). \tag{5}$$

The arguments $x_l$ can be computed, e.g., sequentially in $O(n^4)$ operations, while the polynomial $B_n(x_1, x_2, \ldots, x_n)$ is computed recursively in $O(n^2)$ operations from

$$B_n(x_1, \ldots, x_n) = \sum_{m=1}^{n} \binom{n-1}{m-1} x_m B_{n-m}(x_1, \ldots, x_{n-m})$$

with the initial value $B_0 = 1$. Decomposition (3) using Eq. (5) allows computing the determinant in $O(n^4)$ operations.

*Lemma 2.* Let $C$ be a nonsingular $n \times n$ matrix, then

$$\det(C) C^{-1} = \sum_{s=0}^{n-1} C^s \sum_{k_1, k_2, \ldots, k_{n-1}} \prod_{l=1}^{n-1} (-1)^{k_l+1} \frac{\operatorname{tr}(C^l)^{k_l}}{k_l! l^{k_l}}, \tag{6}$$

where the sum runs over the powers of $C$ and the sets of non-negative integers $(k_1, k_2, \ldots, k_{n-1})$ satisfying the equation

$$s + \sum_{l=1}^{n} l k_l = n - 1. \tag{7}$$

A combinatorial proof of Lemma 2 is given in Ref. [3]. Here we offer an alternative

*Proof.* The inverse of $C$ may be represented as

$$\det(C)(C^{-1})_i^j = \frac{1}{(n-1)!} \varepsilon_{i i_2 \ldots i_n} \varepsilon^{j j_2 \ldots j_n} C_{j_2}^{i_2} \ldots C_{j_n}^{i_n}, \tag{8}$$

This equation allows to write

$$\det(C) C^{-1} = \frac{1}{2\pi i} \oint \frac{dz}{z^n} \det(I + zC)(I + zC)^{-1}$$
$$= \frac{1}{2\pi i} \oint \frac{dz}{z^n} (I + zC)^{-1} \exp(\operatorname{tr} \ln(I + zC)).$$

Expanding the right-hand side in a power series around $z = 0$, one gets

$$\det(C) C^{-1} = \frac{1}{2\pi i} \oint \frac{dz}{z^n} \sum_{s=0}^{n-1} (-1)^s z^s C^s \sum_{k_1, k_2, \ldots, k_{n-1}} z^{k_1 + 2k_2 + \ldots + (n-1)k_{n-1}} \prod_{l=1}^{n-1} (-1)^{lk_l + k_l} \frac{\operatorname{tr}(C^l)^{k_l}}{k_l! l^{k_l}}$$
$$= \sum_{s=0}^{n-1} C^s \sum_{k_1, k_2, \ldots, k_{n-1}} \prod_{l=1}^{n-1} (-1)^{k_l+1} \frac{\operatorname{tr}(C^l)^{k_l}}{k_l! l^{k_l}}.$$

The sum is taken over the powers of $C$ and the sets of non-negative integers $(k_1, k_2, \ldots, k_{n-1})$ satisfying Eq. (7). ∎

*Remark 3.* The right-hand side of Eq. (6) can be presented in terms of the complete Bell polynomials of arguments $x_l = -(l-1)! \operatorname{tr}(C^l)$ as follows

$$\det(C) C^{-1} = \sum_{s=1}^{n} C^{s-1} \frac{(-1)^{n-1}}{(n-s)!} B_{n-s}(x_1, \ldots, x_{n-s}). \tag{9}$$

## 3. Semi-Characteristic Polynomial of Two Skew-Symmetric Matrices

We formulate first an auxiliary result, which represents the analog of Eq. (8).
*Proposition.* Suppose $A$ is a $2n \times 2n$ nonsingular skew-symmetric matrix, then

$$\mathrm{pf}(A)(A^{-1})_{ji} = \frac{1}{2^{n-1}(n-1)!} \varepsilon_{iji_2j_2\ldots i_nj_n} A^{i_2j_2}\ldots A^{i_nj_n}. \tag{10}$$

*Proof.* We start from Cayley's theorem on Pfaffians, which states

$$\mathrm{pf}(A)^2 = \det(A). \tag{11}$$

Taking derivative of Eq. (11), one gets

$$2\mathrm{pf}(A)\frac{\delta \mathrm{pf}(A)}{\delta A^{ij}} = \frac{\delta \det(A)}{\delta A^{ij}} = A^{-1}{}_{ji}\det(A) = A^{-1}{}_{ji}\mathrm{pf}(A)^2,$$

from which we obtain

$$\frac{\delta \mathrm{pf}(A)}{\delta A^{ij}} = \frac{1}{2}(A^{-1})_{ji}\mathrm{pf}(A). \tag{12}$$

This equation is equivalent to Eq. (10). ∎

We also give for Eq. (10) a combinatorial
*Proof.* Multiplying the right-hand side of Eq. (10) by the matrix $A^{kj}$, we obtain

$$A^{kj}\varepsilon_{iji_2j_2\ldots i_nj_n} A^{i_2j_2}\ldots A^{i_nj_n} = \varepsilon_{iji_2j_2\ldots i_nj_n} A^{[kj}A^{i_2j_2}\ldots A^{i_nj_n]} \tag{13}$$

Skew-symmetry in the indices $j, i_2, j_2, \ldots, i_n, j_n$ is available by virtue of the convolution with the Levi-Civita symbol $\varepsilon_{iji_2j_2\ldots i_nj_n}$. On the other hand, the indices $k, j$ have skew-symmetry because of skew-symmetry of the matrix $A^{kj}$. Hence, the tensor in the right-hand side of the equation is skew-symmetric in the whole set $k, j, i_2, j_2, \ldots, i_n, j_n$. Consequently, the right-hand side is proportional to the Levi-Civita symbol $\varepsilon^{kji_2j_2\ldots i_nj_n}$. This fact is highlighted by placing the indices inside the square brackets. We can write

$$A^{[kj}A^{i_2j_2}\ldots A^{i_nj_n]} = \varepsilon^{kji_2j_2\ldots i_nj_n}\frac{1}{(2n)!}\varepsilon_{k'j'i'_2j'_2\ldots i'_nj'_n}A^{k'j'}A^{i'_2j'_2}\ldots A^{i'_nj'_n}.$$

Taking the definition of the Pfaffian into account, one has

$$\frac{1}{(2n)!}\varepsilon_{k'j'i'_2j'_2\ldots i'_nj'_n}A^{k'j'}A^{i'_2j'_2}\ldots A^{i'_nj'_n} = \frac{2^n n!}{(2n)!}\mathrm{pf}(A).$$

The convolution of the Levy-Civita symbols in the $2n-1$ indices results in

$$\varepsilon_{iji_2j_2\ldots i_nj_n}\varepsilon^{kji_2j_2\ldots i_nj_n} = (2n-1)!\delta_i^k,$$

and so

$$A^{kj}\varepsilon_{iji_2j_2\ldots i_nj_n}A^{i_2j_2}\ldots A^{i_nj_n} = (2n-1)!\frac{2^n n!}{(2n)!}\delta_i^k\mathrm{pf}(A).$$

Finally, we obtain

$$A^{kj}\frac{\delta \mathrm{pf}(A)}{\delta A^{ij}} = \frac{1}{2^n(n-1)!}(2n-1)!\frac{2^n n!}{(2n)!}\delta^k{}_i\mathrm{pf}(A) = \frac{1}{2}\delta^k{}_i\mathrm{pf}(A),$$

in agreement with Eq. (12). ∎

The main results of this paper are stated as Lemmas 3 and 4:
*Lemma 3.* Let $A$ and $B$ be $2n \times 2n$ skew-symmetric matrices, then

$$\text{pf}(A)\text{pf}(B) = \sum_{k_1,k_2,\ldots,k_n} \prod_{l=1}^{n} (-1)^{k_l} \frac{\text{tr}((AB)^l)^{k_l}}{k_l! 2^{k_l} l^{k_l}}, \tag{14}$$

where the sum runs over sets of non-negative integers $(k_1, k_2, \ldots, k_n)$ satisfying the equation

$$\sum_{l=1}^{n} l k_l = n. \tag{15}$$

*Proof.* We consider the product

$$\begin{aligned}
\text{pf}(A)\text{pf}(B) &= \frac{1}{2^n n!} \varepsilon_{i_1 j_1 i_2 j_2 \ldots i_n j_n} A^{i_1 j_1} A^{i_2 j_2} \ldots A^{i_n j_n} \frac{1}{2^n n!} \varepsilon^{k_1 l_1 k_2 l_2 \ldots k_n l_n} B_{k_1 l_1} B_{k_2 l_2} \ldots B_{k_n l_n} \\
&= \left(\frac{1}{2^n n!}\right)^2 \sum_{\sigma \in S_{2n}} \text{sgn}(\sigma) A^{i_1 j_1} A^{i_2 j_2} \ldots A^{i_n j_n} B_{\sigma(i_1)\sigma(j_1)} B_{\sigma(i_2)\sigma(j_2)} \ldots B_{\sigma(i_n)\sigma(j_n)}.
\end{aligned} \tag{16}$$

In the sum over permutations, each term is a product of certain number of the traces of products of matrices $A$ and $B$. Let the trace comprising one matrix $A$ occurs $k_1 \geq 0$ times, the trace comprising two matrices $A$ occurs $k_2 \geq 0$ times,..., and the trace comprising $n$ matrices $A$ occurs $k_n \geq 0$ times. Since the left-hand side is the homogeneous function of the degree $n$, the equation (15) holds. In the equations (14) and (15) the values $(k_1, k_2, \ldots, k_n)$ have the identical meanings.

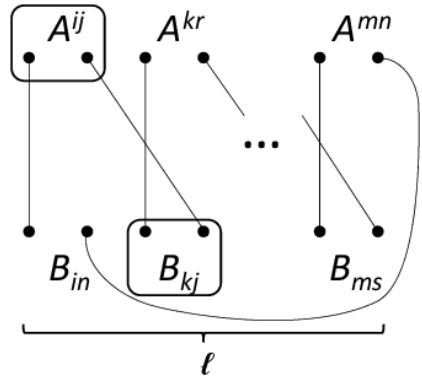

**Figure 1.** *The standard scheme of the contractions in a sample of $l$ matrices A. Parity of the permutation of the standard scheme equals $(-1)^{l+1}$; it is determined by the order of indices of B. Any contraction involving $l$ matrices A reduces to the standard scheme using an even permutation of the upper and lower indices.*

On the right-hand side of Eq. (15), the configuration $(k_1, k_2, \ldots, k_n)$ occurs

$$N = \frac{n!}{k_1! k_2! \ldots k_n!}$$

times. From the set of all terms with this configuration, we select one arbitrary term and consider a trace entering into it, under the sign of which there are $k_l$ matrices $A$ and the same number of matrices $B$. One and the same trace corresponds to the orders of matrices $A$, characterized by their cyclic permutations. Such configurations are identical, but treated so far as independent. Consequently, $N$ should be divided by $l$. Further, the upper indices are contracted with the lower indices, so the results are expressed in terms of the matrix product $AB$. While the set of $A$ is constrained by the configuration $(k_1, k_2, \ldots, k_n)$, the choice of the matrices $B$ is free from constraints. After summing up all possibilities for the contractions of the upper indices with the lower indices, we obtain the factor $n!$ as additional common factor. Further, there exist $2^l$ possibilities for the contractions of the upper indices. By virtue of skew-symmetry of $A$ and the Levi-Civita symbol, these alternatives are summed up with one common sign. The same freedom exists for matrices $B$ with the lower indices, which gives additional factor $2^l$. The very last contraction of two upper and two lower indices has, however, the multiplicity 2 instead of 4, so the common factor $2^l \times 2^l$ is divided by 2. The sign attributed to the configuration follows from Fig. 1, which shows the standard contraction scheme. All other schemes are derivable from the standard one by an even permutation of the indices. Parity of the standard scheme corresponds to a cyclic shift of the lower right indices of $B$ by one unit. It is equal to $(-1)^{l+1}$. Finally, we should take the product of the number of the alternatives for $l = 1, 2, \ldots, n$. We estimated thereby the number of terms on the right-hand side of Eq. (16) with the configuration $(k_1, k_2, \ldots, k_n)$ and found its common sign. The result, consequently, takes the form

$$\mathrm{pf}(A)\mathrm{pf}(B) = \left(\frac{1}{2^n n!}\right)^2 n!n! \sum_{k_1,k_2,\ldots,k_n} \prod_{l=1}^{n} \frac{1}{k_l!} \left(\frac{2^l \times 2^l (-1)^{l+1} \mathrm{tr}((\tilde{B}A)^l)}{2l}\right)^{k_l},$$

which is equivalent to Eq. (14) under the constraint (15). ∎

We also give an independent proof of Lemma 3, based on Cauchy's theorem.

*Proof.* The Pfaffian is the homogeneous function of the degree $n$ of matrix elements. In this connection, the identity follows

$$\mathrm{pf}(A)\mathrm{pf}(B) = \frac{1}{2\pi i} \oint \frac{dz}{z^{n+1}} \mathrm{pf}(\tilde{B}^{-1} + zA)\mathrm{pf}(B). \tag{17}$$

The integration with respect to the complex variable $z$ is carried out along a closed contour counterclockwise in a vicinity of the origin. The following relations are valid up to the sign

$$\mathrm{pf}(A)\mathrm{pf}(B) = \pm \frac{1}{2\pi i} \oint \frac{dz}{z^{n+1}} \left(\det(\tilde{B}^{-1} + zA)\det(\tilde{B})\right)^{1/2}$$

$$= \pm \frac{1}{2\pi i} \oint \frac{dz}{z^{n+1}} \left(\det(I + z\tilde{B}A)\right)^{1/2}$$

$$= \pm \frac{1}{2\pi i} \oint \frac{dz}{z^{n+1}} \exp(\frac{1}{2}\mathrm{tr}(\ln(I + z\tilde{B}A))).$$

Here, $\tilde{B}$ is the transposed matrix $B$. We expand further the exponent in powers of $z$. All the terms of the order $O(z^{n+1})$ can be discarded, because after expansion of the exponential in a series around $z = 0$ such terms vanish after the integration along the contour. The last line gives the determinant in the form of the exponent of a trace. In this way, we obtain

$$\mathrm{pf}(A)\mathrm{pf}(B) = \pm \frac{1}{2\pi i} \oint \frac{dz}{z^{n+1}} \exp(\frac{1}{2} \mathrm{tr} \sum_{l=1}^{n} (-1)^{l+1} \frac{z^l (\tilde{B}A)^l}{l})$$

$$= \pm \frac{1}{2\pi i} \oint \frac{dz}{z^{n+1}} \sum_{k_1,k_2,\ldots,k_n} \prod_{l=1}^{n} \frac{((-1)^{l+1} \frac{z^l}{2l} \mathrm{tr}(\tilde{B}A)^l)^{k_l}}{k_l!}$$

$$= \pm \frac{1}{2\pi i} \oint \frac{dz}{z^{n+1}} \sum_{k_1,k_2,\ldots,k_n} \prod_{l=1}^{n} (-1)^{lk_l + k_l} z^{lk_l} \frac{\mathrm{tr}((\tilde{B}A)^l)^{k_l}}{k_l! 2^{k_l} l^{k_l}} \tag{18}$$

$$= \pm \frac{1}{2\pi i} \oint \frac{dz}{z^{n+1}} \sum_{k_1,k_2,\ldots,k_n} z^{k_1 + 2k_2 + \ldots + nk_n} \prod_{l=1}^{n} (-1)^{k_l} \frac{\mathrm{tr}((\tilde{B}A)^l)^{k_l}}{k_l! 2^{k_l} l^{k_l}}$$

$$= \pm \sum_{k_1,k_2,\ldots,k_n} \prod_{l=1}^{n} (-1)^{k_l+1} \frac{\mathrm{tr}((\tilde{B}A)^l)^{k_l}}{k_l! 2^{k_l} l^{k_l}}.$$

The second line shows the exponent of the sum in the form of product of the exponents, each of which is expanded in the series around $z = 0$. In the next two lines, the terms entering the sums are reordered. From the representation of next to the last line it is clear that the integral is different from zero only in the case of the condition (9) holds true. The last line gives the Pfaffian.

It remains to fix the sign of the right-hand side of the equation. Let $A = B$. In the left-hand side we get $\det(A)$. After transformations, the right-hand side becomes

$$\pm \sum_{k_1,k_2,\ldots,k_n} \prod_{l=1}^{n} (-1)^{k_l} \frac{\mathrm{tr}(A^{2l})^{k_l}}{k_l! 2^{k_l} l^{k_l}} = \pm \sum_{k_1,k_2,\ldots,k_{2n}} \prod_{l=1}^{2n} (-1)^{k_l+1} \frac{\mathrm{tr}(A^l)^{k_l}}{k_l! l^{k_l}}.$$

Here, we make the replacement $2l \to l$ and use the fact that the trace of product of odd number of skew-symmetric matrices is equal to zero. The summation applies to sets of non-negative integers $(k_1, k_2, \ldots, k_n)$ satisfying $\sum_{l=1}^{2n} lk_l = 2n$. The right-hand side takes the form of Eq. (3), therefore, in Eq. (18) we have to choose the positive sign, which is consistent with $\mathrm{pf}(A)\mathrm{pf}(B) = \exp(\frac{1}{2}\mathrm{trln}(\tilde{A}B))$. ∎

*Remark 4.* The above relationship can be illustrated as follows. Any skew-symmetric matrix can be reduced to a block diagonal form by a unitary transformation [12, 13]. Accordingly, we represent $A$ as $A = U\Lambda\tilde{U}$, where $\Lambda = \mathrm{diag}(\lambda_1 i\sigma^2, \lambda_2 i\sigma^2, \ldots, \lambda_n i\sigma^2)$. By including a common phase factor of $U$ in the definition of $\lambda_i$, $U$ can taken to be special unitary matrix. The identity $\mathrm{pf}(U\Lambda\tilde{U}) = \det(U)\mathrm{pf}(\Lambda)$ holds for all the matrices $U$, $\Lambda$, and so $\mathrm{pf}(A) = \lambda_1 \lambda_2 \ldots \lambda_n$. Identities $\mathrm{trln}(AB) = \mathrm{trln}A + \mathrm{trln}B$, $\mathrm{trln}(AB) = \mathrm{trln}(U\Lambda\tilde{U}B) = \mathrm{trln}(\Lambda B) + \mathrm{trln}(U) + \mathrm{trln}(\tilde{U})$ are obvious. Since $\det(U) = 1$, one has $\mathrm{trln}(U) = \mathrm{trln}(\tilde{U}) = 0$. The result of the transformations is $\mathrm{trln}(AB) = \mathrm{trln}(\Lambda B)$.

Through similar considerations the matrix $B$ can also be transformed. Let $B = U'\Lambda'\tilde{U}'$, where $\Lambda' = \text{diag}(\lambda_1'i\sigma^2, \lambda_2'i\sigma^2,\ldots, \lambda_n'i\sigma^2)$, then $\text{trln}(\tilde{A}B) = \text{trln}(-\Lambda\Lambda') = 2\ln(\lambda_1\lambda_1') + 2\ln(\lambda_2\lambda_2') +\ldots + 2\ln(\lambda_n\lambda_n')$. Therefore, $\text{pf}(A)\text{pf}(B)$ coincides with $\exp(\frac{1}{2}\text{trln}(\tilde{A}B))$.

*Remark 5.* We have checked with the use of MAPLE that Eq. (14) holds for $B = \text{diag}(i\sigma^2,\ldots, i\sigma^2)$ and all skew-symmetric matrices $A$ of the dimensions 4×4, 6×6, 8×8, and 10×10.

*Remark 6.* The right-hand side of Eq. (14) can be recognized, up to a coefficient, as the complete Bell polynomial of $n$ arguments $x_l = -\frac{1}{2}(l-1)!\,\text{tr}((AB)^l)$:

$$\text{pf}(A)\text{pf}(B) = \frac{1}{n!}B_n(x_1,\ldots,x_n). \tag{19}$$

The Pfaffian is computed from the definition with $(2n)!/(2^n n!)$ complexity. The recursion for $B_n(x_1, x_2,\ldots, x_n)$ allows computing the Pfaffian in $O(n^4)$ operations (cf. Remark 2). The generalized Gaussian elimination computes the Pfaffian in $O(n^3)$ operations.

*Corollary 1.* Let $B = A^{-1}$. A simple calculation based on the relationship (14) gives

$$\begin{aligned}\text{pf}(A)\text{pf}(A^{-1}) &= \sum_{k_1,k_2,\ldots,k_n}\prod_{l=1}^{n}(-1)^{k_l}\frac{n^{k_l}}{k_l!\,l^{k_l}}\\ &= \frac{1}{2\pi i}\oint\frac{dz}{z^{n+1}}\sum_{k_1,k_2,\ldots,k_n}\prod_{l=1}^{n}(-1)^{k_l}\frac{z^{lk_l}n^{k_l}}{k_l!\,l^{k_l}}\\ &= \frac{1}{2\pi i}\oint\frac{dz}{z^{n+1}}\exp\left(-n\sum_{l=1}^{\infty}\frac{z^l}{l}\right) = \frac{1}{2\pi i}\oint\frac{dz}{z^{n+1}}(1-z)^n = (-1)^n.\end{aligned} \tag{20}$$

*Example.* Let $A = i\sigma^2$, where $\sigma^2$ is the 2-nd Pauli matrix, then $A^{-1} = -i\sigma^2$, $\text{pf}(i\sigma^2) = 1$, $\text{pf}(A)\text{pf}(A^{-1}) = -1$, in agreement with the above result.

*Example.* Let $A = \text{diag}(i\sigma^2,\ldots, i\sigma^2)$ is a $2n\times 2n$ matrix. It is not difficult to see that $A^{-1} = -A$, consequently, $\text{pf}(A) = 1$, $\text{pf}(A^{-1}) = (-1)^n$, $\text{pf}(A)\text{pf}(A^{-1}) = (-1)^n$, in agreement with Eq. (20).

*Corollary 2.* Suppose $A$, $B$, and $C$ are $2n\times 2n$ skew-symmetric matrices and $AB + BA = 0$ and $BC + CB = 0$. In such a case, the products $AB$ and $BC$ also are skew-symmetric. The conditions of Lemma 3 are thus satisfied with the pairs $AB$, $C$ and $A$, $BC$. Equation (14) then gives $\text{pf}(AB)\text{pf}(C) = \text{pf}(A)\text{pf}(BC)$.

*Example.* Let $A = \gamma^1$, $B = \gamma^1\gamma^3 = \text{diag}(i\sigma^2, i\sigma^2)$, and $C = \gamma^3$, where $\gamma^j$ are the Dirac $\gamma$-matrices in the standard representation. The matrices $A$, $B$, and $C$ fit the conditions of Corollary 2. It is not difficult to see that $\text{pf}(A) = \text{pf}(B) = \text{pf}(C) = \text{pf}(AB) = \text{pf}(BC) = 1$, so the equation $\text{pf}(AB)\text{pf}(C) = \text{pf}(A)\text{pf}(BC)$ is satisfied.

*Lemma 4.* Let $A$ and $B$ be $2n\times 2n$ skew-symmetric matrices, then

$$\text{pf}(A)\text{pf}(B)A^{-1} = -\sum_{s=0}^{n-1}(BA)^s B\sum_{k_1,k_2,\ldots,k_{n-1}}\prod_{l=1}^{n-1}(-1)^{k_l}\frac{\text{tr}((BA)^l)^{k_l}}{k_l!\,2^{k_l}l^{k_l}}, \tag{21}$$

where the sum runs over the powers of product $BA$ and sets of non-negative integers $(k_1, k_2,\ldots, k_{n-1})$ satisfying the equation

$$s + \sum_{l=1}^{n-1}lk_l = n-1. \tag{22}$$

*Proof.* By virtue of Eq. (10) the left-hand side of the equation is the homogeneous function of degree $n-1$ with respect to the elements of $A$. We thus write the left-hand side in the form of a contour integral:

$$\begin{aligned}\text{pf}(A)\text{pf}(B)A^{-1} &= \frac{1}{2\pi i}\oint\frac{dz}{z^n}\text{pf}(B^{-1}+zA)\text{pf}(B)(B^{-1}+zA)^{-1}\\ &= \frac{1}{2\pi i}\oint\frac{dz}{z^n}\text{pf}(B^{-1}+zA)\text{pf}(B)\sum_{s=0}^{n-1}(-z)^s(BA)^s B.\end{aligned}$$

In the second line, we expand $(B^{-1} + zA)^{-1}$ in a power series of $z$. The terms $O(z^n)$ give vanishing contributions to the integral, so the series is truncated. The product of the Pfaffians can be transformed as follows:

$$\begin{aligned}\text{pf}(A)\text{pf}(B)A^{-1} &= \pm\frac{1}{2\pi i}\oint\frac{dz}{z^n}\sum_{s=0}^{n-1}(-z)^s(BA)^s B\exp\left(\frac{1}{2}\text{trln}(I+zBA)\right)\\ &= \pm\frac{1}{2\pi i}\oint\frac{dz}{z^n}\sum_{s=0}^{n-1}(-z)^s(BA)^s B\exp\left(\frac{1}{2}\sum_{l=1}^{n-1}\frac{(-1)^{l+1}z^l}{l}\text{tr}(BA)^l\right)\end{aligned}$$

$$= \pm \frac{1}{2\pi i}\oint \frac{dz}{z^n}\sum_{s=0}^{n-1}(-z)^s (BA)^s B \sum_{k_1,k_2,\ldots,k_{n-1}}\prod_{l=1}^{n-1}(-1)^{k_l+lk_l} z^{lk_l} \frac{(\mathrm{tr}((BA)^l))^{k_l}}{k_l!2^{k_l} l^{k_l}}$$

$$= \pm \sum_{s=0}^{n-1}(BA)^s B \sum_{k_1,k_2,\ldots,k_{n-1}}\prod_{l=1}^{n-1}(-1)^{k_l+1}\frac{(\mathrm{tr}((BA)^l))^{k_l}}{k_l!2^{k_l} l^{k_l}}.$$

In order to fix the sign of the right-hand side, we set $A = B$. One has

$$\det(A) A^{-1} = \pm \sum_{s=0}^{n-1} A^{2s+1} \sum_{k_1,k_2,\ldots,k_{n-1}}\prod_{l=1}^{n-1}(-1)^{k_l+1}\frac{(\mathrm{tr}(A^{2l}))^{k_l}}{k_l!(2l)^{k_l}}$$

$$= \pm (-1)^n \sum_{s=0}^{2n-1} A^s \sum_{k_1,k_2,\ldots,k_{2n-1}}\prod_{l=1}^{2n-1}(-1)^{k_l+1}\frac{(\mathrm{tr}(A^l))^{k_l}}{k_l! l^{k_l}},$$

where the summation over the non-negative integers $(k_1, k_2,\ldots, k_{2n-1})$ is constrained by the condition

$$s + \sum_{l=1}^{2n-1} l k_l = n-1.$$

In the transition to the second line, the replacement $2l \to l$ is made; the summation over $s$ runs from zero to $2n-1$; $l$ runs from one to $2n-1$. If for some odd $l$ the number $k_l$ is nonzero, the trace of the product of $l$ matrices $A$, and therefore, the whole contribution to the sum, vanishes. Consequently, $k_l$ can be set equal to zero for all odd values of $l$. The second line of the equation according to Lemma 2 is an expansion in powers of the inverse matrix $A$. This implies that for the right-hand side of the equation the sign $(-1)^n$ must be selected, in accordance with the equation (21). ∎

*Remark 7.* We have checked with the use of MAPLE that Eq. (21) holds for $B = \mathrm{diag}(i\sigma^2,\ldots,i\sigma^2)$ and all skew-symmetric matrices $A$ of the dimensions 4×4, 6×6, 8×8, and 10×10.

*Remark 8.* The right-hand side of Eq. (21) can be presented in terms of the complete Bell polynomials of arguments $x_l = -\frac{1}{2}(l-1)!\,\mathrm{tr}((BA)^l)$ as follows

$$\mathrm{pf}(A)\mathrm{pf}(B) A^{-1} = -\sum_{s=1}^{n}(BA)^{s-1} B \frac{1}{(n-s)!} B_{n-s}(x_1,\ldots,x_{n-s}). \tag{23}$$

*Example.* Let $B = A^{-1}$. In such a case, according to Corollary 1, $\mathrm{pf}(A)\mathrm{pf}(B) A^{-1} = (-1)^n A^{-1}$. From other hand, the direct use of Eq. (21) gives

$$\mathrm{pf}(A)\mathrm{pf}(B) A^{-1} = -\sum_{s=0}^{n-1} A^{-1} \sum_{k_1,k_2,\ldots,k_{n-1}} \prod_{l=1}^{n-1}(-1)^{k_l}\frac{(2n)^{k_l}}{k_l!2^{k_l} l^{k_l}}$$

$$= -\sum_{s=0}^{n-1} A^{-1} \frac{1}{2\pi i}\oint \frac{dz}{z^n} z^s \exp\left(-n\sum_{l=1}^{\infty}\frac{z^l}{l}\right)$$

$$= -\sum_{s=0}^{n-1} A^{-1} \frac{1}{2\pi i}\oint \frac{dz}{z^n} z^s \exp(n\ln(1-z))$$

$$= -A^{-1} \frac{1}{2\pi i}\oint \frac{dz}{z^n} \sum_{s=0}^{n-1} z^s (1-z)^n$$

$$= -A^{-1} \frac{1}{2\pi i}\oint \frac{dz}{z^n} \frac{1-z^n}{1-z}(1-z)^n = -A^{-1} \frac{1}{2\pi i}\oint \frac{dz}{z^n}(1-z)^{n-1} = (-1)^n A^{-1}.$$

We consider that $\mathrm{tr}((A^{-1}A)^l) = 2n$. The second line restricts the permissible sets of the $(k_1, k_2,\ldots, k_{n-1})$. The contour integral selects sets, which satisfy Eq. (22). The summation over the $(k_1, k_2,\ldots, k_{n-1})$ in the integrand can be extended from zero to infinity, given that the powers of $z$ higher than $n-1$ do not contribute to the integral. The series is then summed up easily. The subsequent transformation is obvious. Thus, another check shows that the overall sign in Eq. (21) is correct.

*Corollary 3.* Equation (21) can be written in the symmetric form with regard to the matrices $A$ and $B$:

$$\mathrm{pf}(A)\mathrm{pf}(B)(AB)^{-1} = -\sum_{s=0}^{n-1}(AB)^s \sum_{k_1,k_2,\ldots,k_{n-1}}\prod_{l=1}^{n-1}(-1)^{k_l}\frac{\mathrm{tr}((AB)^l)^{k_l}}{k_l!2^{k_l} l^{k_l}}. \tag{24}$$

If a $2n \times 2n$ nonsingular matrix $C$ admits the representation as the product of two $2n \times 2n$ skew-symmetric matrices $A$ and $B$, the decomposition of $C^{-1}$ in powers of $C = AB$ requires the $n$ terms, whereas the decomposition of Lemma 2 requires the $2n$ terms in general.

*Definition.* Let $A$ and $B$ are $2n \times 2n$ skew-symmetric matrices and $\det(B) \neq 0$. Semi-characteristic polynomial of $A$ and $B$ is the polynomial of the form

$$p_n(\lambda) = \text{pf}(A - \lambda B^{-1})\text{pf}(B).$$

The pairs of skew-symmetric matrices obey a counterpart of the Cayley-Hamilton theorem:

*Corollary 4.* Let $A$ and $B$ are $2n \times 2n$ skew-symmetric matrices. Combining Eqs. (14) and (24), we obtain the matrix identity

$$\sum_{s=0}^{n}(AB)^s \frac{1}{(n-s)!} B_{n-s}(x_1,\ldots,x_{n-s}) = 0, \qquad (25)$$

where $x_l = -\frac{1}{2}(l-1)!\,\text{tr}((BA)^l)$. There exists a $2n \times 2n$ matrix $U$ that brings $AB$ into its diagonal form: $UABU^{-1} = \text{diag}(\lambda_1,\ldots,\lambda_{2n})$. Applying this transformation to Eq. (25), we find that the eigenvalues of $AB$ are zeros of the semi-characteristic polynomial

$$p_n(\lambda) = \sum_{s=0}^{n} \lambda^s \frac{1}{(n-s)!} B_{n-s}(x_1,\ldots,x_{n-s}). \qquad (26)$$

There exist at most $n$ distinct roots of the equation $p_n(\lambda) = 0$, so the set of $2n$ eigenvalues $\lambda_l$ contains at most $n$ distinct numbers. This statement also is a consequence of the characteristic equation $\det(AB - \lambda I) = 0$, because $\det(AB - \lambda I) = (\text{pf}(A - \lambda B^{-1})\,\text{pf}(B))^2 = p_n(\lambda)^2$. Zeros of the characteristic polynomial $\det(AB - \lambda I)$ are therefore of the second order (or higher). The relashionship $p_n(\lambda) = \text{pf}(A - \lambda B^{-1})\text{pf}(B)$ can easily be proved starting from the representation

$$\text{pf}(A - \lambda B^{-1})\text{pf}(B) = \sum_{s=0}^{n} \lambda^s \frac{1}{2\pi i} \oint \frac{dz}{z^{s+1}} \text{pf}(A - zB^{-1})\text{pf}(B)$$

$$= \sum_{s=0}^{n} \lambda^s \frac{1}{2\pi i} \oint \frac{dz}{z^{s+1}} z^n \exp(\frac{1}{2}\text{tr}\ln(1 - \frac{1}{z}AB))$$

and proceeding in the same way as in the derivation of Eq. (14). $p_n(\lambda)$ is the minimal polynomial of $AB$ provided its zeros are simple.

## 4. Conclusion

The eigenvalues of a $n \times n$ matrix $C$ may be computed by searching for roots of the characteristic equation $\det(C - \lambda I) = 0$. The polynomial coefficients are computed efficiently using the Faddeev – LeVerrier recursion, from which it follows, in particular, that these coefficients depend on the traces of powers of $C$. The polynomial coefficients are given finally by the complete Bell polynomials of the arguments related in a simple manner to the traces of powers of $C$.

The product of skew-symmetric matrices, $A$ and $B$, has a specificity which allows to terminate the power series in $AB$ earlier than prescribed by the Cayley-Hamilton theorem. We operate, accordingly, with the quasi-characteristic polynomial $p_n(\lambda) = \text{pf}(A - \lambda B^{-1})\text{pf}(B)$, whose degree, $n$, is two times lower that the degree $2n$ of the charactericsic polynomial of $AB$. In this paper, we derived the corresponding trace identities for the Pfaffian, the inverse of $A$ and $B$, and proved the analog of the Cayley-Hamilton theorem: $p_n(AB) = 0$. These results are of particular interest for matrices with multi-dimensional indices, e.g., spin, isospin, spatial, and others related to the various symmetries. The multiplication operation and the trace function do not mix the matrix sub-indices, so the symmetries can be controlled using the trace identities at all stages of computing the determinant, the Pfaffian, and inverse matrices.

## Acknowledgments

This work was supported in part by RFBR Grant No. 16-02-01104 and Grant No. HLP-2015-18 of the Heisenberg-Landau Program.

## References

[1] N. Bourbaki, Elements of mathematics, 2. Linear and multilinear algebra (Addison-Wesley, 1973).

[2] F. R. Gantmacher, The Theory of Matrices. Vol. 1 and 2. (Chelsea Pub. Co., N. Y. 1959).

[3] L. A. Kondratyuk, M. I. Krivoruchenko, Superconducting quark matter in SU (2) color group, Z. Phys. A 344, 99 (1992).

[4] L. S. Brown, Quantum Field Theory (Cambridge University Press, 1994).

[5] Q.-L. Hu, Z.-C. Gao, and Y. S. Chen, Matrix elements of one-body and two-body operators between arbitrary HFB multi-quasiparticle states, Phys. Lett. B 734, 162 (2014).

[6] Yang Sun, Projection techniques to approach the nuclear many-body problem, Phys. Scr. 91, 043005 (2016).

[7] Wei Li, Heping Zhang, Dimer statistics of honeycomb lattices on Klein bottle, Möbius strip and cylinder, Physica A 391, 3833 (2012).

[8] G. H. Hardy and S. Ramanujan, Asymptotic formulae in combinatory analysis, Proc. London Math. Soc. 17, 75 (1918).

[9] Ya. V. Uspensky, Asymptotic expressions of numerical functions occurring in problems concerning the partition of numbers into summands, Bull. Acad. Sci. de Russie 14, 199 (1920).

[10] M. I. Krivoruchenko, Recurrence relations for the number of solutions of a class of Diophantine equations, Rom. J. Phys. 58, 1408 (2013).

[11] E. T. Bell, Partition polynomials, Ann. Math. 29, 38 (1928).

[12] Th. Voronov, Pfaffian, in: Concise Encyclopedia of Supersymmetry and Noncommutative Structures in Mathematics and Physics, Eds. S. Duplij, W. Siegel, J. Bagger (Springer, Berlin, N. Y. 2005), p. 298.

[13] D. C. Youla, A normal form for a matrix under the unitary congruence group, Canad. J. Math. 13, 694 (1961).